\documentclass[12pt]{article}
\usepackage{amssymb}
\usepackage{epsfig}

\newcommand{\bq}{\begin{equation}}
\newcommand{\eq}{\end{equation}}
\newcommand{\bqa}{\begin{eqnarray}}
\newcommand{\eqa}{\end{eqnarray}}
\newcommand{\ben}{\begin{enumerate}}
\newcommand{\een}{\end{enumerate}}
\newcommand{\bc}{\begin{center}}
\newcommand{\ec}{\end{center}}
\newcommand{\bqb}{\begin{eqnarray*}}
\newcommand{\eqb}{\end{eqnarray*}}

\def\gsim{\gtrsim}
\def\lsim{\lesssim}
 
\def\pr#1#2#3{ Phys. Rev. ${\bf{#1}}$ (#2) #3}
\def\prl#1#2#3{ Phys. Rev. Lett. ${\bf{#1}}$ (#2) #3}
\def\pl#1#2#3{ Phys. Lett. ${\bf{#1}}$ (#2) #3}

\def\np#1#2#3{ Nucl. Phys. ${\bf{#1}}$ (#2) #3}
\def\zp#1#2#3{ Z. f. Phys. ${\bf{#1}}$ (#2) #3}
\def\ijmp#1#2#3{ Int. J. Mod. Phys. ${\bf{#1}}$ (#2) #3}

\def\ie{{\it i.e.\/}}
\def\eg{{\it e.g.\/}}

\def\etal{{\it et.al.\/}}

\global\nulldelimiterspace = 0pt

\def\wtil#1{\widetilde{#1}}
\def\ol#1{\overline{#1}}

\def\L{ {\cal L }}
\def\O{ {\cal O }}
\def\M{ {\cal M }}

\def\lamU{\Lambda_{U}}

\begin{document}
\pagenumbering{arabic}
\thispagestyle{empty}
\def\thefootnote{\fnsymbol{footnote}}
\setcounter{footnote}{1}
 
\begin{flushright}
PM/97-46 \\ THES-TP 97/10 \\
 hep-ph/9711399 \\
November 1997 \\
revised version
 \end{flushright}
\vspace{2cm}
\begin{center}
{\Large\bf New Physics Signatures in Dijets at Hadron 
Colliders}\footnote{Partially supported by the EC contract 
CHRX-CT94-0579.}
 \vspace{1.5cm}  \\
{\large G.J. Gounaris$^a$, D.T. Papadamou$^a$ and F.M. Renard$^b$}
\vspace {0.5cm}  \\

$^a$Department of Theoretical Physics, University of Thessaloniki,\\
Gr-54006, Thessaloniki, Greece.\\
\vspace{0.2cm}
$^b$Physique
Math\'{e}matique et Th\'{e}orique,
UPRES-A 5032\\
Universit\'{e} Montpellier II,
 F-34095 Montpellier Cedex 5.\\

\vspace {1cm}
 
{\bf Abstract}
\end{center}
We show how to detect and disentangle at the upgraded 
Tevatron and  at LHC,
the effects of the three
purely gluonic $dim=6$ $SU(3)\times
SU(2) \times U(1)$ CP-conserving and CP-violating
gauge invariant operators 
$\ol{\O}_{DG}$, $\O_G$ and $\wtil{\O}_{G}$. These operators
are inevitably generated by New Physics (NP), if the heavy particles 
responsible for it are coloured. 
We establish the relations between their coupling
constants and the corresponding NP scales defined through the
unitarity relations. We then study the sensitivity and limits 
obtainable through production processes involving one or two jets,
and express these limits in terms of
the NP scales implied by unitarity. A detailed comparison with the
results of the studies of the analogous electroweak operators,
is also made.\\

 PACS: 12.38.Qk, 12.60.-i, 13.85.-t, 14.70.-e .\\

\vspace{0.5cm}
\def\thefootnote{\arabic{footnote}}
\setcounter{footnote}{0}
\clearpage

\section{Introduction}

The main experimental indication for 
the physics beyond the Standard Model (SM) is up to now only 
provided by non-accelerator observations, like the
apparent predominance of the dark matter in the Universe and 
the non-vanishing of the neutrino masses. On the contrary,
accelerator experiments at LEP and lower energies seem to be 
quite consistent with SM \cite{SM1, SM2}; except of an    
admittedly meager evidence for an enhancement 
compared to the QCD predictions, 
in deep inelastic $e^+p$ scattering  at HERA \cite{H1,
ZEUS}, and the apparent excess of jet production at large
transverse energy $E_T$, recently observed by the CDF
Collaboration at the Tevatron \cite{CDF-ET, Schellman}. 

These mean that any  acceptable form of New Physics (NP), 
involving only enhancements or depressions of the number of events 
with respect to the standard QCD expectations,
can at most only 
induce very slight modifications of SM.
Of course it is almost certain that  such NP effects 
cannot always have an unambiguous
interpretation. Therefore, the search of such NP effects 
requires that, for any
hint of an experimental signal, various possible interpretations
are explored and  ways of discriminating among them 
are identified. \par

One possible way to search for NP,
is  within the scenario that  no new particles, 
apart from one scalar
standard-like Higgs, will be reachable in the future colliders.
Under such a philosophy,  
a classification of a rather wide class of NP models has been
presented in \cite{Buchmuller, top-op, top-Young}. 
At the energy range of the foreseeable colliders, NP is then assumed
to be described by the set of all $dim=6$  $SU(3)\times
SU(2) \times U(1)$ gauge invariant operators 
involving an isodoublet Higgs field, the quarks of the third
family,  and of course the 
gauge bosons, inevitably introduced by the gauge principle whenever
a derivative appears. The complete list of the CP conserving
such operators has been given in \cite{top-op, top-Young},
while the CP violating ones have appeared in 
\cite{top-CP, Tsirigoti-CP}. \par

Most of these operators, are either directly or indirectly constrained 
by LEP1 or lower energy experiments
\cite{LEP1-constraints}, or will be constrained by the direct
production of weak gauge boson, Higgs and top or $b$-quark 
at $e^-e^+$ and $\gamma \gamma $ colliders \cite{LEP2, NLC}, as well as 
at hadron colliders \cite{LHC, top-Young, top-CP}. There exist
two CP conserving operators though \cite{Simmons0},
which cannot be efficiently constrained through the processes
just listed, namely
\bq
\overline{\O}_{DG}
  = 2 ~ (D_{\mu} \overrightarrow G^{\mu
\rho}) (D^{\nu} \overrightarrow G_{\nu \rho}) \ \ \
  , \ \ \label{ODG}
\eq

\noindent
and
\bq
\O_G =  {1\over3!}~  f_{ijk}~ G^{i\mu\nu}
  G^{j}_{\nu\lambda} G^{k\lambda}_{\ \ \ \mu} \ \ \
 .  \ \ \   \label{OG}   
\eq
These operators will  inevitably appear  
 at  present energies, if the heavy new particles 
generating     NP 
are coloured \cite{Tsirigoti, top-op, Papadopoulos}.
The second one, $\O_G$, is a genuine purely gluonic operators;
while the equations of motion of the QCD lagrangian implies 
that $\overline{\O}_{DG}$ is
essentially equivalent to a four quark operator
\cite{Simmons0}.
They are on the same footing as the operators 
$\ol{\O}_{DW}$ and $\O_W$, responsible for inducing 
anomalous electroweak triple gauge couplings \cite{LEP2, NLC}.  
 \par

In addition to these two CP conserving operators, 
there exist also the unique analogous CP violating $dim=6$ operator  
\cite{Weinberg, Tsirigoti-CP}  
\bq  
\wtil{\O}_G =  {1\over3!}~  f_{ijk}~ \wtil{G}^{i\mu\nu}
  G^{j}_{\nu\lambda} G^{k\lambda}_{\ \ \ \mu} \ \ \
 ,  \ \ \   \label{OGtilde}   
\eq 
where $\wtil{G}^i_{\mu \nu}=
1/2\epsilon_{\mu\nu\lambda\sigma}G^{i\lambda \sigma}$.
$\wtil{\O}_G$  caused considerable discussion some time ago,
which started from Weinberg's observation that the existing 
limits on the neutron electric dipole moment, (combined with
reasonable assumptions on how to calculate such a low
energy parameter), put an extremely strong constraint on
the $\wtil{\O}_G$ coupling. Since indirect constraints are always 
submitted to ambiguities, it is worthwhile to also
have a direct check of this operator at the Tevatron and LHC. \par

In order to complete the study of the
aforementioned set of the $dim=6$  $SU(3)\times
SU(2) \times U(1)$ gauge invariant operators, the sensitivity
limits to the three operators $\ol{\O}_{DG}$, $\O_G$ and 
$\wtil{\O}_G$ must also be investigated.
The best way to do this, is to look at single and two jet
production  at hadron colliders.  
Such studies have already appeared in 
\cite{Simmons0, Zeppenfeld},  while an application to
the Tevatron was performed in \cite{Simmons}. The identification
of the NP operators was subsequently also  
discussed in $Z\to 4~jets$ \cite{Duff}, 
in 3 jet production \cite{Dixon},
and more recently in top-antitop production \cite{Simmonst}.
We should also quote \cite{Papadopoulos} who study the 
dijet angular distribution, but choose to concentrate on a specific 
coherent combination of four CP-conserving $dim=6$ operators,
two of which coincide with $\ol{\O}_{DG}$ and $\O_G$.\par

In the present work we consider in detail  the implications
of  $\ol{\O}_{DG}$, $\O_G$ and $\wtil{\O}_{G}$, 
for single and dijet production at the Tevatron and LHC.  
Thus, in Section 2 we first establish 
the unitarity constraints on their  couplings, which 
allows us to check the self-consistency of our
perturbative treatment, and  to obtain an unambiguous connection
between the sensitivity limit on the 
value of each NP coupling, and the NP scale at which the
corresponding operator might be generated. 
In this same Section 2 we also present the observables 
studied, while the
related technical details are put  in the Appendix.
In Section 3,  we  present the  
effects of each of  the three operators on the 
inclusive single jet and the dijet production at the Tevatron (present and
upgraded stages), and at the
LHC.  We compute the rapidity, transverse energy, invariant mass
and angular distributions and we examine how they reflect the presence
of NP contributions. We then establish the sensitivity limits
in terms of couplings constants and new physics scales 
for each operator separately.
In Section 4 we discuss ways to disentangle
contributions from the three operators, and give special illustrations
using the dijet angular distribution.
Finally in Section 5 we  compare with the results 
previously obtained for the
analogous electroweak operators involving $W$ bosons, and we draw
the conclusion on the picture of NP which should come out of 
the whole set of $dim=6$  $SU(3)\times
SU(2) \times U(1)$ gauge invariant operators.

\section{Formalism.}

The NP effects are described through the effective
Lagrangian
\bq
 \L_{NP} ~= ~f_{DG} \ol{\O}_{DG}+
~f_{G} g_s \O_{G}+ ~\wtil{f}_{G} g_s \wtil{\O}_{G} \ \ ,
\label{NPlagrangian}
\eq
where the dimensional NP couplings $f_j$
are conveniently  measured  in $TeV^{-2}$.\par

\subsection{Unitarity constraints.}

We first establish the relations between these dimensional coupling
constants and the scales at which unitarity is saturated,
following the same method as in \cite{unitarity, top-op} and
working separately for each operator. From the
strongest partial wave unitarity constraint, we derive a relation
between the dimensional coupling $f_j$ and the energy at which unitarity
is saturated, which is denoted as $\lamU$. It can be
interpreted as a practical definition of the NP scale; (elsewhere
denoted by $\Lambda_{NP}$ \cite{unitarity, top-op}). 
This procedure allows us to 
unambiguously associate to each value of the NP coupling $f_j$, a
corresponding value of the scale $\lamU$.\par

Thus, for $\ol{\O}_{DG}$, the strongest constraint arises from the 
two flavour- and colour-singlet $J=0$ channels 
$|q\bar q \pm \pm \rangle $ and is given by
\bq
|f_{DG}| ~=~\left ( \frac{3}{8\alpha_s} \right
) \frac{1}{\lamU^2}\simeq \frac{3.7}{\lamU^2} 
\ \ \ \ . \label{DGunitarity}
\eq\par

For $\O_G$, the strongest constraint arises from the  
colour-singlet $J=0$ channels $|gg \pm \pm \rangle $ and it is
given by
\bqa
f_G & = & \left (\frac{10}{\sqrt {g_s}}+
g_s \right ) \frac{1}{\lamU^2}\simeq \frac{10.4}{\lamU^2}  \
\ \ \ \ \mbox{ for} ~f_G > 0 \ , \nonumber \\  
f_G & = & \left ( -\, \frac{10}{\sqrt{g_s}}+
g_s \right ) \frac{1}{\lamU^2}\simeq -\, \frac{ 8 }{\lamU^2}  \
\ \ \ \ \mbox{ for} ~f_G < 0 \ . \label{Gunitarity}
\eqa
The same channels also give the strongest constrain for
the $\wtil{\O}_G$ operator, which now leads  to
\bq
|\wtil{f}_G|  =  \left (\frac{10}{\sqrt{g_s}} \right ) 
\frac{1}{\lamU^2}\simeq  \frac{9.5}{\lamU^2}  \
\ \ \ .  \ \label{tildeGunitarity} 
\eq\par 

 These relations allow us to check the
self-consistency of our perturbative results, in the sense that
we can   check that the values of the NP couplings used, are
such that the associated $\lamU$ scale is larger than the 
effective subprocess energy. In other words, $\hat s \lsim \lamU^2$ is
a necessary condition in order to guaranty that the description of NP in
terms of $dim=6$ operators is valid, for the process considered.\par

\subsection{Jet production at hadron colliders.}

The observables  at a hadron collider that we consider are
the inclusive one jet $pp  \to  j \ ...$, and the 
exclusive\footnote{As far as the number of jets is concerned.}
 two jet production  $pp  \to  i\ j \ ...$ production, where 
$i$, $j$ denote the  light quark or gluon jet
produced. 
In the single jet study we consider the distribution
\bq
\frac{d\sigma(pp \to  j \ ...)}{d\eta dx_T} \ \ ,
\eq
\noindent
where $(\eta, ~E_T)$ are the pseudorapidity 
and transverse energy of the observed jet.
While, in the two jet case, we discuss the invariant mass $M_{ij}$ 
and angular $\chi$ distribution
\bq
\frac{d\sigma(pp  \to  i\ j \ ...)}{dM_{jj}^2  d\chi }
\eq
where
$\chi=(1+|\cos \theta^*|)/(1-|\cos \theta^*|)$, and $\theta^*$
is  the subprocess scattering angle in the parton c.m. frame.
The  relevant formulae expressing these distributions in terms of parton
structure functions and subprocess cross sections, are given in Appendix
A.\par

\section{Results}

{\bf a) General features.}

The NP contributions to the squared amplitudes of the available 
subprocesses with massless partons, 
averaged (summed) over the initial (final)
spins and colours, are given for completeness in Table 1,
where we only keep terms up to quadratic in the NP couplings.
The results for $\ol{\O}_{DG}$, $\O_G$ 
have been first derived in \cite{Simmons0, Simmons}, 
where it has also been observed that the $\O_G$ contribution
does not interfere with the QCD amplitudes for massless partons;
while for the $\ol{\O}_{DG}$ case a non vanishing interference
with QCD exists. This is reflected by the fact that in the
results in Table 1, there exist linear contributions in $f_{DG}$, but only
quadratic in $f_G$.  Because of this, 
the operator $\ol{\O}_{DG}$ produces strong effects on 
quark-(anti)quark subprocesses at high 
energies\footnote{This is essentially due to the fact that
it induces a kind of renormalization of the gluon propagator,
which is so strong that it cancels its  $1/q^2$ behaviour, and
forces the four quarks to interact locally.} 
and consequently the sensitivity to its
coupling $f_{DG}$ is quite large. 
Such a  greater sensitivity  is also confirmed by the value of the
corresponding unitarity scale, which is
found to be much higher than the average energy of the 
subprocess,  (see below).\par

The operators $\O_G$ and $\wtil{\O}_G$, 
give identical contributions to  the spin averaged (summed) 
squared amplitudes for  all $2\to 2$ subprocess
involving massless partons; see Table 1. 
For both operators, but for different basic reasons (in the $\O_G$
case due to the helicity dependence \cite{Simmons0}, and in the
$\wtil{\O}_G$ one due to the CP violation),  
there is no interference between NP and the 
QCD amplitudes, so that the dominant
contributions only arise at the quadratic level,
$f_G^2$ or $\wtil{f}_G^2$ level.
Such contributions are in general
on the same footing as the contributions induced by possible 
$dim=8$ operators, which happen to have a non vanishing
interference with the standard QCD 
results\footnote{Contributions linear to $f_G$ first appear in 
three jet effects; see end of Section 3 below \cite{Dixon}.}. 
As a consequence the sensitivity is weaker than in the 
$\ol{\O}_{DG}$ case. Thus, if some indications of a non
vanishing contribution from these operators is found, then a
detail program for disentangling them from possible $dim=8$
contributions should be pursued.
The disentangling of these two operators among themselves also
requires a difficult study of
CP-odd observables. We will come back to this point in the
conclusion. \par

In spite of these difficulties, we have found that observable
effects due to the operators $\O_G$ or  $\wtil{\O}_G$,
can be obtained for values of the coupling constants
for which our perturbative treatment is acceptable; \ie\@ for  NP 
couplings which are below (but close to) the unitarity limit for all
relevant subprocess energies. Thus, acceptable 
values of the couplings exist, for which the effect of $\O_G$ 
or  $\wtil{\O}_G$ on single jet production is similar to the one due
to $\ol{\O}_{DG}$. This raises the problem of disentangling
the contributions arising from the 
$\ol{\O}_{DG}$ on the one hand, and the sum of the
squares of the couplings of the operators  
$\O_G$ and $\wtil{\O}_{G}$ on the other hand. We  discuss it
in the next section and in the conclusion.\par

{\bf b) Applications to Tevatron and LHC.}

We next turn to the discussion of the results presented in  
Fig.1. For any physical quantity, we always plot the 
relative variation induced by NP with respect to the QCD
prediction. Such a presentation is very convenient since 
it can be adequately calculated using the leading log QCD formalism. 
In Fig.1a we  give the relative NP effect for the
$E_T$ distribution at the Tevatron. 
The results are averaged over $\eta$ in the region
$0.1 < |\eta|< 0.7$, and the
CDF \cite{CDF-ET} and the D0 \cite{D0-ET}
data are also shown \cite{Albrow}. 
In Fig.1b the same results are given for LHC. \par

As mentioned already, the $\O_G$ ($\wtil{\O}_G$) prediction is
independent of the sign of the NP coupling  and it always
enhances the QCD expectation. 
On the contrary, if the source of 
NP is $\ol{\O}_{DG}$, then an enhancement appears essentially
only for negative $f_{DG}$ values. In the examples given in the
figures, the magnitudes of the
$f_G$, $f_{DG}$ (for $f_{DG} <0$ ) were chosen, so
that they induce effects of similar magnitudes for the
$E_T$ distributions.
For the Tevatron case in particular, the NP effects are chosen
so that they are roughly consistent with an NP enhancement 
like the one tolerated  by the CDF data. 
In Fig.1a,b we also give the scales $\lamU$ where the
NP interactions would saturate 
unitarity. We have checked that  the relevant
subenergies of all
partonic processes, are sufficiently below the
corresponding $\lamU$ scales, so that our perturbative 
treatment is safe.
Thus, an attempt to understand
an NP effect of \eg\@ the order suggested by the Tevatron data
in terms of $\ol{\O}_{DG}$, necessitates couplings very much below the
unitarity limit. On the contrary, if we try to understand the
same kind  of effect in terms of $\O_G$ (or $\wtil{\O}_G$), 
we need an NP coupling close to (but still below) the unitarity 
limit. \par

To appreciate the sensitivity of the future hadron colliders 
to the $E_T$ distribution shown in Fig.1a,b,  we
define the signal $S$ and background $B$ 
by the number of events for 
\bq
S =(NP+QCD)^2 -(QCD)^2 \ \ \ , \ \ B=(QCD)^2 \  \ \ . \
\label{S/B}
\eq
For the
upgraded $2~TeV$ Tevatron with  luminosity
$\L=10^4 pb^{-1}$,  we consider the  events in the range 
$0.1 < |\eta| < 0.7$ and $0.06 < E_T < 0.65$, which then give  
\bqa
\ol{\O}_{DG} \  & \Longrightarrow & \frac{S}{\sqrt B} =  \{
-3~  f_{DG} +0.31 f_{DG}^2 \} \sqrt{\L} \ \ \ , 
\label{ET-TEV-sens-DG} \\ 
\O_G \ , \ \wtil{\O}_G  & \Longrightarrow  & \frac{S}{\sqrt B} = 0.011  
f_G^2  \sqrt{\L} \ \  ,
\label{ET-TEV-sens-G}
\eqa
where the NP couplings are measured in $TeV^{-2}$, and we 
only keep terms up to
quadratic in the NP couplings. The meaning of 
(\ref{ET-TEV-sens-G}) is that an NP signal at the upgraded 
Tevatron,  like the one shown by the curves 
in Fig.1a for  $f_{DG}=-0.3$, would imply an enhancement of about 
$S/\sqrt B =92$ standard deviations, with respect to
the QCD expectations. 
Similarly, for the $\O_G$ ($\wtil{\O}_G$) case
with $f_G=\pm 6$ ($\wtil{f}_G=\pm 6$), the results would imply 
effects of
$S/\sqrt B=39.6$ standard deviations.
This suggests that the upgraded Tevatron could
improve present  Tevatron sensitivity limits by roughly an order of
magnitude.\par

Correspondingly for LHC, integrating over all events in the
range $0.1 \leq |\eta|\leq 0.7 $ and  $0.1 <E_T < 3.5~ TeV$, we get
\bqa
\ol{\O}_{DG} \  & \Longrightarrow & \frac{S}{\sqrt B} = \{
-11.2~ f_{DG} +4.16 f_{DG}^2  \}\sqrt{\L} \ \ \ , 
\label{ET-LHC-sens-DG} \\ 
\O_G \  & \Longrightarrow  & \frac{S}{\sqrt B} = 0.44  
f_G^2  \sqrt{\L} \ \  .
\label{ET-LHC-sens-G}
\eqa
Eqs. (\ref{ET-LHC-sens-DG}, \ref{ET-LHC-sens-G}) indicate that
LHC  should be sensitive to 
effects of the order of magnitude shown in Fig.1b.
This is inferred from the fact that using $f_{DG}=-0.005$
in  (\ref{ET-LHC-sens-DG}), gives 
$S/\sqrt{B}=5$; while from  $f_G=\pm 0.2$ or $\wtil{f}_G=\pm
0.2$  we get $S/\sqrt B=1.8$. \par 

 Summarizing the $E_T$ discussion we conclude 
that, if we assume that NP observability
demands  $S/\sqrt B \gsim 10$, then an enhancement due to an
$\ol{\O}_{DG}$ contribution will be observable at the upgraded
Tevatron, if $-f_{DG} \gsim 0.03~TeV^{-2}$
($\lamU \lsim 11~TeV$). Correspondingly at LHC, $-f_{DG}$  can
go down to $0.009~TeV^{-2}$, which means that the highest
$\ol{\O}_{DG}$  scale 
to which LHC is sensitive, is $\lamU \simeq 20~TeV$. 
In both cases, the $\ol{\O}_{DG}$ couplings are 
much below the unitarity limits for the suitable average energies
of the various subprocesses. We also note that in this case, 
if $f_{DG}>0$, then  we could even end up in a situation where
no NP enhancement is actually realized at the present Tevatron with
$\sqrt s =1.8 ~TeV$, while at LHC we might even have a
depression; see Figs.1a,b. \par

For the $\O_G$ and $\wtil{\O}_G$
cases on the other hand, the 
corresponding sensitivity limits to the NP coupling  
are at $f_G \simeq 3.0~TeV^{-2}$ ($\lamU \simeq 1.7~TeV$) 
for the upgraded Tevatron,
and  $0.48~TeV^{-2}$ ($\lamU \simeq 4.3~TeV$)  for LHC. 
These limits  (arising from quadratic NP
contributions) are in fact as large as they
can possibly be, in consistency with our perturbative
treatment.\par 

Further information on the $\O_G$  couplings could also be
obtained \cite{Dixon} by looking at 3-jet final states, in which
the  $\O_G$ contribution is enhanced by its 
appearance  at the linear level, but at the same time it is suppressed 
by an extra power of $\alpha_s$. In \cite{Dixon} an application was
made for the Tevatron. It was shown that in the domain where 2 jets are
collinear, the azimuthal distribution ($cos2\phi$) is especially 
sensitive to the $\O_G$ contribution. Observability limits taking
into account all experimental constraints have not yet been
obtained using this method, but we feel that
they are more or less of the same size as those in the dijet
studies, for which the expected
sensitivity limits lie close to the unitarity bounds.
The 2-jet and 3-jet final states give therefore complementary
information and should both be studied at future machines
in the search of NP interactions. \par

\section{Disentangling of the three operators}

We have seen above that the NP effects in the $E_T$
distribution, induced by  
$\ol{\O}_{DG}$ for $f_{DG}<0$, could be  very similar to those
implied by $\O_G$ or $\wtil{\O}_G$. 
In \cite{Simmons0} it was already noticed that the angular
distribution (see the t and u dependence in Table 1) should allow 
distinguishing  $\ol{\O}_{DG}$ from $\O_G$. 
The interest in the angular distribution is further enhanced by
remarking that it depends on different features of the
 structure functions than the $E_T$ distribution, and
can therefore be used to test whether a possible $E_T$
effect really necessitates an NP explanation \cite{Albrow}.
To study this in more detail, we present in Fig.2 the relative 
variation  to  the $\chi$ distribution induced 
by NP. Notice, that if the NP contribution had exactly
the same $\chi$ shape as in QCD, then the NP curves in this
figure, would had been completely flat.
In order to increase the NP sensitivity
as much as possible, we have selected the highest range of dijet
invariant mass, compatible with the requirement to have an 
acceptable number of events.
In Fig.2 we use  the same values of the
NP couplings as in Fig.1 and we also give the CDF \cite{CDF-chi}
and D0 \cite{D0-chi} data. We also note that because of
$M_{jj}=E_T(\chi+1)/\sqrt{\chi}$, we get $E_T\gsim 80~GeV$,
for all points in Fig.2; which means that the chosen NP couplings are
always below the unitarity limit and the perturbative 
treatment is  safe.\par

According to Figs.2a,b,  
 $\ol{\O}_{DG}$ for  $f_{DG}<0 $,  
and $(\O_G$, $\wtil{\O}_G)$, always enhance the number of
central events; with the $(\O_G$, $\wtil{\O}_G)$ enhancement
being the strongest one. On the other hand for positive $f_{DG}$ (\eg\@
$f_{DG} = 0.3~TeV^{-2}$), we could even have a depression in the number
of events at the upgraded Tevatron and LHC, (with the depression
being somewhat stronger in the central region); while no
effect would actually arise at the present Tevatron 
with  $\sqrt s =1.8~TeV$. \par   

The difference of the $\chi$-shapes does not look observable at
the present Tevatron because of the large errors of existing 
data; compare Figs.2a,b. 
But at the upgraded Tevatron and the LHC, because of the larger
energy and the expected
luminosity $\L=10^4pb^{-1}$, the statistical errors should
decrease sufficiently to allow discrimination of the two shapes.
This is suggested by the comparison of  Figs.2c,d with 
Table 2, where the signal over
background ratio (see (\ref{S/B})) is given for the indicated 
$\chi$-bins and sufficiently high dijet mass. 
Table 2 implies that for 
the upgraded Tevatron and  the $\chi$
bins indicated there, we get   
$S/\sqrt B = (48.7,~~21,~~13)$  for $f_{DG}=-0.3$,
and  $S/\sqrt B = (39.6,~~9,~~2.8)$  for $f_G=\pm 6$ 
(or $\wtil{f}_G=\pm 6)$. This difference in the $\chi$ shape,
combined with the similarity of the $E_T$ distributions, should
allow disentangling between $\ol{\O}_{DG}$ and ($\O_G$,
$\wtil{\O}_G$), at least for couplings of the order of those
indicated in the figures. In the LHC case, the results in Table
2  for all events in the  range $1< \chi <5 $, 
give $S/\sqrt B =10.6$ for $f_{DG}=-0.005$, and  $S/\sqrt B = 15.5$ for  
$f_G=\pm 0.2$ or $\wtil{f}_G=\pm 0.2$. Thus, in the predictions
shown in Fig.2d, there is a $\sim 5$
standard deviation difference between the $\ol{\O}_{DG}$ and the 
$(\O_G,~\wtil{\O}_G)$ results for $1<\chi <5$, and essentially
no difference for higher $\chi$. Therefore, it should 
be possible to discriminate between these two alternatives at
LHC, at the level of the NP couplings used in the figure.  \par

A useful quantity for describing the dependence of the
$\chi$-shape on the dijet mass, is  
$R_{\chi}$ defined by the ratio \cite{CDF-chi, D0-chi}
\bq
R_\chi ~ = ~ \frac{N(1 < \chi < \chi_1)}{N(\chi_1 < \chi
< \chi_{max})} \ \ .  \
\eq
In Fig. 3 we plot the relative variation of $R_\chi$, with
respect to the QCD prediction. The same NP couplings
are used, as in the previous figures. The results in Figs.3a
correspond to the present  Tevatron and are compared with
the CDF data using $\chi_1=2.5$, $\chi_{max}=5$  \cite{CDF-chi};
while in Fig.3b the D0 data for  four experimental points
with $\chi_1=4$ and $(\chi_{max}=20,~~20,~~13,~~11)$ 
\cite{D0-chi}, are compared with the corresponding NP
prediction. Finally Fig.3c,d examples of NP effects are given
using the same couplings as before. Concerning the CDF data, it
is amusing to remark that although the  $\O_G, ~\wtil{\O}_G$ 
prediction for the $\chi$ distribution in Fig.2a is not
particularly close to these data, the $R_\chi$ prediction 
seems to follow their central values. This
simply emphasizes again the point made in the introduction
that the discovery of NP through studies of enhancement or
depression effects,
necessitates that all possible interpretations and ways of
analysing the data should be tried. \par

An independent way of disentangling the effects of $\O_G$ was
recently discussed in \cite{Simmonst} through top+antitop quark
production. For heavy quarks, a contribution to the cross section
linear in $f_G$ and proportional to $m^2_t$ appears. It is for
example found
that this operator contributes  more strongly to the transverse
momentum distribution, than the other
gluonic operators and the  $dim=6$ $SU(3) \times SU(2) \times U(1)$ gauge
invariant operators involving directly the top quark. So this
leads us to expect that a global study of light quark and gluon jets 
and heavy quark production might allow a good disentangling
of the various operators.

\section{Conclusion}
In this paper we have studied the 3 purely gluonic operators
$\ol{\O}_{DG}$ and $\O_G$, $\wtil{\O}_G$. We have followed 
the same method that we used for the other 
$dim=6$ $SU(3) \times SU(2) \times U(1)$ gauge
invariant operators involving
gauge bosons, Higgs bosons and heavy quarks \cite{NLC1}.\par 
We have first established the
unitarity constraints relating the coupling constants to the energy
scale at which unitarity is saturated. This allows  to
associate unambiguously to the coupling constant of each operator, 
an effective NP scale where this operator is generated. 
We have then proceeded to a phenomenological
analysis of the effects of these operators in hadron collisions at
present and future machines. More precisely we have concentrated on
single jet and dijet production at the Tevatron and at the LHC.
On the basis of these, we determined the observability limits 
for each operator, which we describe below in terms of the 
NP scale $\lamU$ that can be possibly probed.\par

The operator $\ol{\O}_{DG}$ (whose effect is essentially a
renormalization of the gluon propagator) produces strong effects on
quark (anti)-quark subprocesses. The sensitivity limits 
correspond to an NP scale
\bq 
\lamU=11,~~~20 ~~~TeV
\eq
at the upgraded Tevatron and LHC respectively.
This operator is the gluonic analogue of the operator $\ol{\O}_{DW}$,
which also produces direct modifications of the W boson propagator and
which is already strongly constrained at LEP1/SLC by $\lamU
\simeq 10~ TeV$,  and should be further studied at LEP2 and NLC 
with the expected sensitivities of
$17$ and $38~ TeV$ respectively, see \cite{DW}.\par

The two other operators, $\O_G$ and its CP-violating partner 
$\wtil{\O}_G$, lead to identical effects in all two-body processes
involving massless unpolarized quarks and gluons. They contribute 
through diagrams
involving their anomalous three-gluon coupling. Due to a different
helicity dependence, the part of the amplitude which is
linear in this anomalous term does not interfere with the SM amplitude.
The anomalous contribution to the cross section only starts with
quadratic terms \cite{Simmons0}. The sensitivity to these processes 
in then weaker than the one
to $\ol{\O}_{DG}$. In terms of NP scales one now finds
\bq 
\lamU=1.7,~~~4.3 ~~~TeV \ \
\eq
for the upgraded Tevatron and LHC respectively.
These operators are the analogues of the operators
$\O_W$ and $\wtil{\O}_W$ which have been extensively studied in
$e^+e^-\to W^+W^-$ and $\gamma \gamma \to W^+W^-$, and whose 
effects are also much less constrained than
the $\ol{\O}_{DW}$ ones\footnote{The reason is that they
contribute only indirectly at Z peak.}. Thus,
the expected sensitivity scales for $\O_W$ and $\wtil{\O}_W$
are  $1.5~TeV$ at LEP2 and $10 ~TeV$ at NLC, see \cite{W}. \par

So there is an interesting parallelism between the situation for gluonic
and for electroweak operators.\par

For disentangling  the
effects of the three gluonic operators, it is difficult to pursue the
analogy with W-boson operators. In the electroweak case, 
the disentangling of $\ol{\O}_{DW}$ from $\O_W$ and
$\wtil{\O}_W$ is obvious, because only $\ol{\O}_{DW}$
 contributes directly to 4-fermion processes, whereas the
other two operators contribute directly to W pair production. 
For the gluonic operators the situation is
the same, replacing fermions by quarks and W bosons by gluons; but it
is now very difficult (or even impossible for light quarks), 
to identify the respective
contributions of quarks and gluons in the
initial and  final states. In the case of heavy quarks $\ol{\O}_{DG}$ 
contributes to $q\bar q\to t\bar t$  but not to $gg \to t\bar t$,
whereas $\O_G$ and and $\wtil{\O}_G$ contribute to 
$gg \to t\bar t$ but not to $q\bar q\to t\bar t$; but the game becomes
more complex because genuine operators involving directly the top quark
may contribute \cite{Simmonst}.\par 

In the present paper we have found that the disentangling of
$\ol{\O}_{DG}$ effects from those of  $\O_G$ and $\wtil{\O}_G$ can be
achieved by looking at the dijet angular distribution. The
$\ol{\O}_{DG}$ contribution is closer to the SM one (particularly 
for $f_{DG} >0$), whereas  the $\O_G$ and $\wtil{\O}_G$
contributions are more
concentrated in the central region. The use of the parameter
$R_{\chi}$, essentially the ratio of the number of
events in the central and in the peripheral regions, 
allows to clearly separate  these two types of distributions.
This study however requires an important luminosity which will only be
available at upgraded Tevatron and LHC.\par
The disentangling of the CP-conserving $\O_G$
from the CP-violating $\wtil{\O}_G$ is much more difficult. 
In the $W$ case,  methods were proposed
using asymmetries in $W^+$ and $W^-$ decay distributions in the
process $e^+e^-\to W^+W^-$  \cite{GRS}, and 
$\gamma \gamma \to W^+W^-$ \cite{Choi} for polarized initial states. 
There is no analogue possibility here. The only solution seems
to be a direct study of 
CP violation in multijet production processes. One should then extend
the analysis
done in \cite{Dixon} and consider CP-odd observables in order to reach
the effect due to $\wtil{\O}_G$, for example
along the lines of the study done for $Z$ decay into three jets
\cite{CP}.  \par

When all these informations will be available from experiments, one
should have at our disposal a clear panorama of residual NP effects
expressed in terms of $dim=6$ $SU(3) \times SU(2) \times U(1)$ gauge
invariant operators. We have emphasized the parallelism between the
operators
involving gluons and the ones involving $W$ bosons. This should tell
us about the role of colour in the underlying NP dynamics.
More generally, a comparison of the possible effects or of the
upper limits in the various sectors (electroweak gauge sector, Higgs
sector, heavy quark sector, gluonic sector) should give valuable
informations on the nature and the origin of the NP dynamics.
\par 

\vspace*{0.5cm}
\noindent
\underline{Acknowledgments}\\
We like to thank Jean-Loic Kneur for a helpful  discussion.

\vspace*{2cm}
 
\renewcommand{\theequation}{A.\arabic{equation}}
\setcounter{equation}{0}
\setcounter{section}{0}

{\large \bf Appendix A: Kinematics for one jet and two jet studies}

When the hadrons $A$ and
$B$ collide  inelasticly through their respective partons 
$a$ and $b$, the transverse energy $E_T$ and rapidity
distributions of an inclusively produced  
jet $j$ associated with the light parton $c$,  are  expressed as 
\cite{Stirling}
\bqa
\frac{d\sigma(AB \to  j \ ...)}{d\eta dx_T} & = &
\frac{\pi e^\eta \alpha_s(\mu)^2}{s} \cdot \nonumber \\
& &
\sum_{abcd} \int_{x_{1min}}^1\frac{dx_1}{x_1^2} f_{a/A}(x_1,
\mu) f_{b/B}(x_2,\mu)\ol{|\M(ab\to cd)|^2}
\frac{1}{1+\delta_{cd}} \ \ .
\label{ET}
\eqa
Here,  all parton masses are neglected, 
\bq
x_T= \frac{2E_T}{\sqrt s} \ \ \ , 
\ \ \ x_{1min}=\frac{x_T e^\eta}{2-x_T e^{-\eta}}
\ \ \ , \ \ x_2=\frac{x_1 x_T e^{-\eta}}{2 x_1- x_Te^\eta} \ ,
\label{xET}
\eq
$(\eta, ~E_T)$ are  the pseudorapidity and transverse energy
of the observed jet, and the QCD scale $\mu$ is usually taken
equal to  $E_T/2$, or  $E_T$. In (\ref{ET}),
\bq
F(ab\to cd) \equiv g_s^2 \M(ab \to cd) \ \ 
\eq  
denotes the invariant amplitude of the partonic subprocess
$(ab\to cd)$. $\ol{|\M(ab\to cd)|^2}$ is the squared reduced amplitude 
averaged (summed) over the initial (final) spins and colours 
 and the Mandelstam parameters are written as   
\bq
\hat s= s x_1x_2 \ \ \ , \ \ \ \hat t = -x_1 \sqrt{s}E_T
e^{-\eta} \ \ \ \hat u=- \hat s - \hat t \ \ ,
\label{MandelstamET}
\eq 
where $s$ is the c.m. energy-squared of the Collider.\par

The   $\chi\equiv (1+|\cos \theta^*|)/(1-|\cos \theta^*|)$
and invariant mass $M_{ij}^2\equiv \hat s$
distributions for the dijet are  similarly  given by 
\cite{Stirling} 
\bqa
\frac{d\sigma(AB \to  i\ j \ ...)}{dM_{ij}^2  d\chi } & = &
\frac{2\pi \alpha_s^2(\mu) }{ s M_{ij}^2 (1+\chi)^2} \cdot \nonumber \\
& &
\sum_{abcd} \int_{-\bar \eta_{lim}}^{\bar \eta_{lim}}
 d\bar \eta f_{a/A}(x_1,
\mu) f_{b/B}(x_2,\mu)\ol{|\M(ab\to cd)|^2}
\frac{1}{1+\delta_{cd}} \ ,
\label{chi}
\eqa
where $\bar \eta$ is the rapidity of the c.m. of the pair of the
two produced jets and
\bq
x_1= \frac{M_{jj}}{\sqrt{s}}e^{\bar \eta} \ \ , \ \  
x_2= \frac{M_{jj}}{\sqrt{s}}e^{-\bar \eta} \ \ ,
\eq
\bq
|\bar \eta |~ \leq ~  \eta_{lim} ~ \equiv ~ min \Bigg \{ ln \left
(\frac{\sqrt{s}}{M_{jj}} \right )~,~ 
\eta_c -\frac{1}{2} ln \chi \Bigg \} \ \ . \ \label{etabarlim}
\eq 
In (\ref{etabarlim}), $\eta_c$ denotes the imposed cut in the
pseudorapidities of each of the two jets; \ie\@
$|\eta_1| \leq \eta_c$, $|\eta_2| \leq \eta_c$.
In all numerical applications here we take $\eta_c=2$.
\par


\newpage

{ Table 1:   $\ol{\O}_{DG}$ and $\O_G$, $\wtil{\O}_G$ 
corrections to $\sum\ol{|F|^2}/g^4$ for the
indicated parton subprocesses with massless partons. 
The colour
and spin indices are
 averaged (summed) over final (initial) states; $n_f$ is the number of
final light quark flavours. }
\begin{center}
\begin{tabular}{|c|c|}
\hline \hline 
\multicolumn{1}{|c|}{} &
\multicolumn{1}{|c|}{}
\\[0.0cm]
\multicolumn{1}{|c|}{}&
\multicolumn{1}{|c|}{$\ol{\O}_{DG}$ }
\\[0.1cm] \hline
$qq \to qq$ & $-\left (\frac{8 \hat s f_{DG} }{27\hat t\hat u} \right )
 \left (\hat s^2+9\hat t^2+9\hat u^2\right )+ 
\left (\frac{32 f_{DG}^2}{9 }\right ) 
\left (\frac{4\hat s^2}{3} +\hat u^2+ \hat t^2 \right )$
\\[0.5cm] 
$q\  q^\prime(\bar q^\prime) \to q\  q^\prime (\bar q^\prime) $ 
&  $-\left (\frac{16\hat s f_{DG} }
{9\hat u\hat t }\right ) \left (\hat s^2+3\hat t^2+3\hat u^2\right )+ 
\left ( \frac{64 f_{DG}^2}{9 }\right ) 
\left (2\hat s^2 +\hat u^2+ \hat t^2 \right )$ \\[0.5cm] 
$q\bar q \to q\bar q $&  $ \left (\frac{32 f_{DG}}{9 }\right )
\left (- \frac{2\hat s^3}{\hat t\hat u}+3\hat s +
\frac{\hat u^4+\hat t^4}{3\hat s\hat t\hat u} + 
2 n_f\, \frac{\hat t^2+\hat u^2}{\hat s} \right )
+ \left ( \frac{64 f_{DG}^2}{9}\right ) 
\left [ 2\hat s^2 +\frac{6 n_f+1}{3}\, (\hat u^2+ \hat t^2)  \right ]$ 
\\[0.5cm] \hline \hline
\multicolumn{1}{|c|}{} &
\multicolumn{1}{|c|}{}
\\[0.0cm]
\multicolumn{1}{|c|}{}&
\multicolumn{1}{|c|}{$\O_G$ or $\wtil{\O}_G$ }
\\[0.1cm] \hline
$ q \bar q \to gg $ & $ \frac{4}{3}\hat u\hat t f_G^2 $\\[0.5cm]
$g\  q (\bar q) \to g\ q (\bar q ) $ &  
$ -  \frac{1}{2}\hat s\hat u f_G^2 $\\[0.5cm]
$ gg \to q \bar q $ & $ \frac{3}{16}\hat u\hat t f_G^2$\\[0.5cm]
$gg \to gg $ & $ \frac{171}{32}\, 
[\hat s^2 -\hat t\hat u] f_G^2 $ \\ \hline
\end{tabular}
\end{center}

\vspace*{0.3cm} 
\begin{center}
{ Table 2: $S/\sqrt B$ ratio for various $\chi$ bins; 
($\L=10^4 pb^{-1}$).}\\
\vspace*{0.3cm}
\begin{tabular}{|c|c|c|}
\hline \hline 
\multicolumn{3}{|c|}{Upgraded Tevatron, 
($\sqrt s =2TeV$), $M_{jj} > 0.4TeV$} 
\\[0.0cm] \hline  
\multicolumn{1}{|c|}{$\chi$}&
\multicolumn{1}{|c|}{$\ol{\O}_{DG}$ } &
\multicolumn{1}{|c|}{$\O_G,~ \wtil{\O}_G $}
\\[0.1cm] \hline
$1 < \chi  < 1.5 $ & $(-1.25 f_{DG} +1.24 f_{DG}^2 )\sqrt \L $ & 
$0.011 f_G^2 \sqrt \L $ \\[0.5cm]
$4.5 < \chi  < 5 $ & $(-0.62 f_{DG} +0.28 f_{DG}^2 )\sqrt \L $ & 
$0.0025 f_G^2 \sqrt \L $ \\[0.5cm]
$9.5 < \chi  < 10 $ & $(-0.4 f_{DG} +0.095 f_{DG}^2 )\sqrt \L $ & 
$0.00079 f_G^2 \sqrt \L $ \\ 
\hline \hline 
\multicolumn{3}{|c|}{LHC ,  $M_{jj} > 3 TeV$} 
\\[0.0cm] \hline  
\multicolumn{1}{|c|}{$\chi$}&
\multicolumn{1}{|c|}{$\ol{\O}_{DG}$ } &
\multicolumn{1}{|c|}{$\O_G,~ \wtil{\O}_G $}
\\[0.1cm] \hline
$1 < \chi  < 1.5 $ & $(-9 f_{DG} +151 f_{DG}^2 )\sqrt \L $ & 
$1.95 f_G^2 \sqrt \L $ \\[0.5cm]
$4.5 < \chi  < 5 $ & $(-3.59 f_{DG} +30.53 f_{DG}^2 )\sqrt \L $ & 
$0.48 f_G^2 \sqrt \L $ \\[0.5cm]
$9.5 < \chi  < 10 $ & $(-1.98 f_{DG} +9.4 f_{DG}^2 )\sqrt \L $ & 
$0.15 f_G^2 \sqrt \L $ \\[0.5cm]
$1 < \chi  < 5 $ & $(-19.8 f_{DG} +303 f_{DG}^2 )\sqrt \L $ & 
$3.9 f_G^2 \sqrt \L $ \\  \hline 
\end{tabular}
\end{center}


\clearpage
\newpage

\begin{figure}[p]
\vspace*{-4cm}
\begin{center}
\mbox{
\epsfig{file=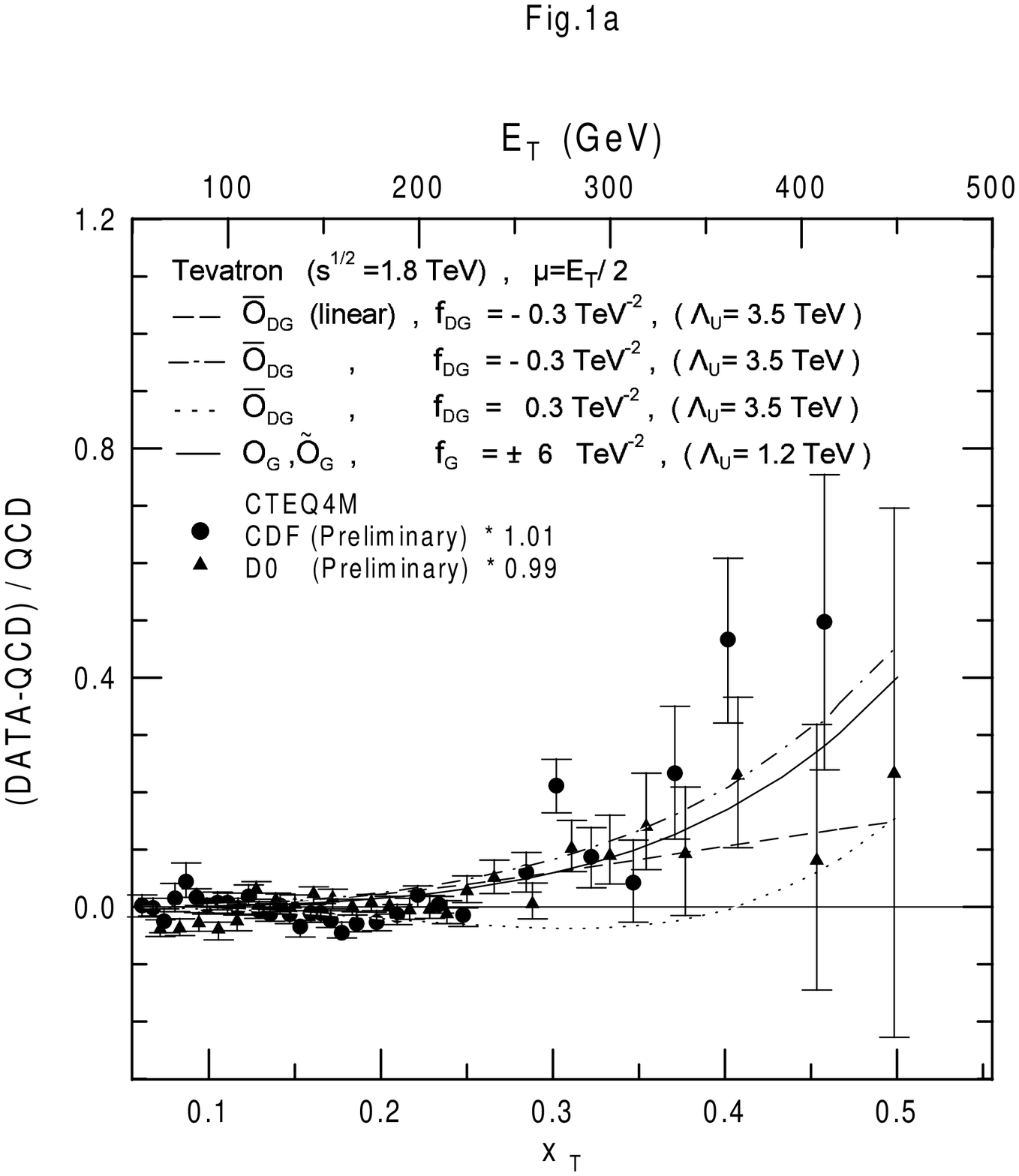,height=12cm}}
\vspace*{-3.7cm}
\end{center}
\hspace{5.5cm} (a)\\
\vspace*{0.5cm}
\begin{center}
\mbox{
\epsfig{file=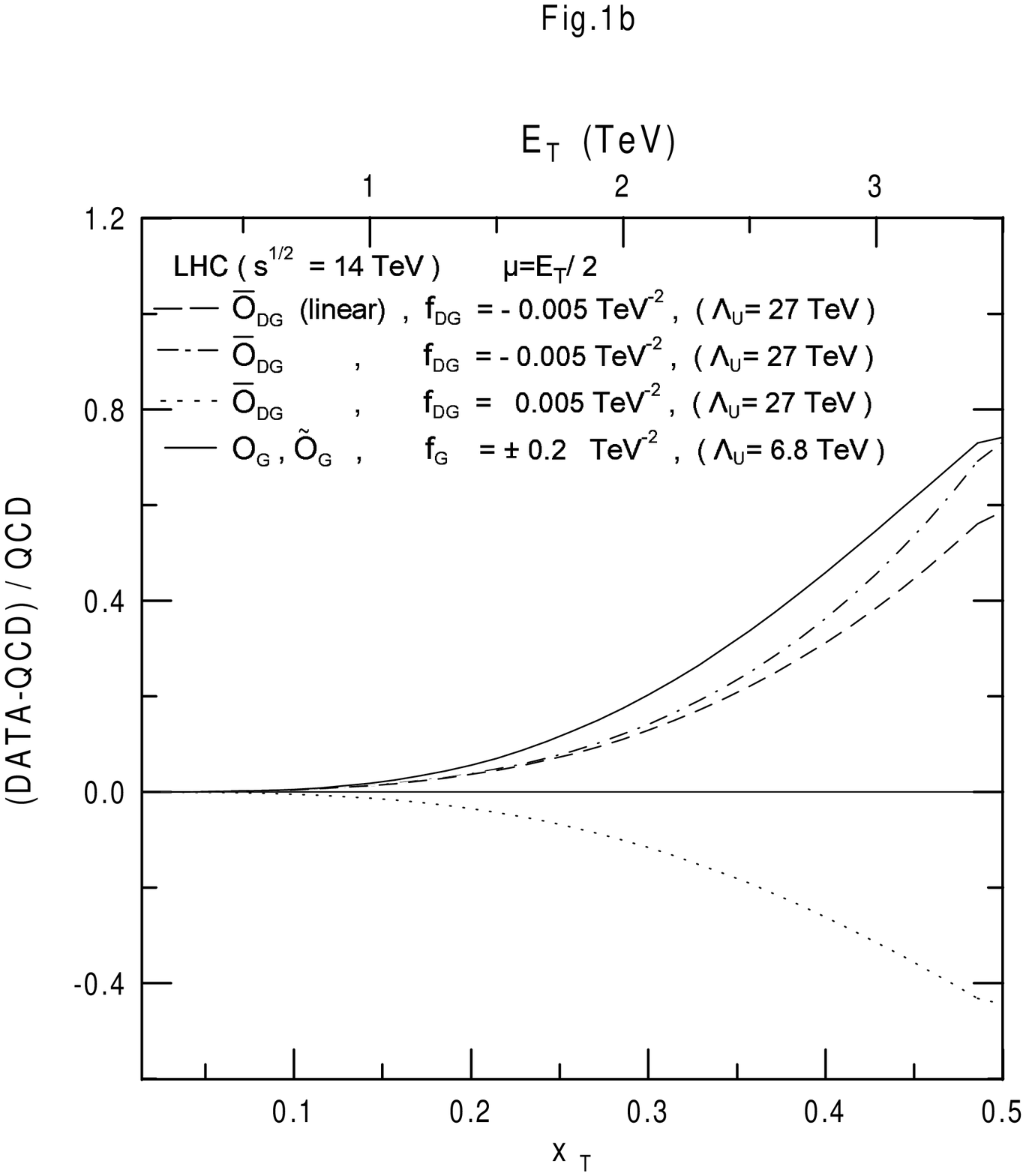,height=12cm}}
\vspace*{-3.7cm}
\end{center}
\hspace{5.5cm} (b)\\
\vspace*{2cm}
\caption[1]{ Possible  NP contribution to the $E_T$ distribution
for the inclusive jet
production averaged over $0.1 \leq |\eta| \leq 0.7 $,
from $\ol{\O}_{DG}$ or $\O_G,~\wtil{\O}_G $;  
compared with the CDF and D0 data at the Tevatron 
\cite{CDF-ET, D0-ET, Albrow} (a),
and a possible LHC signal (b). In the $\wtil{\O}_G$ case the
results correspond to $\wtil{f}_G$ equal to the
indicated $f_G$.} 
\label{et}
\end{figure}

\clearpage
\newpage

\begin{figure}[p]
\vspace*{-5.cm}
\[
\epsfig{file=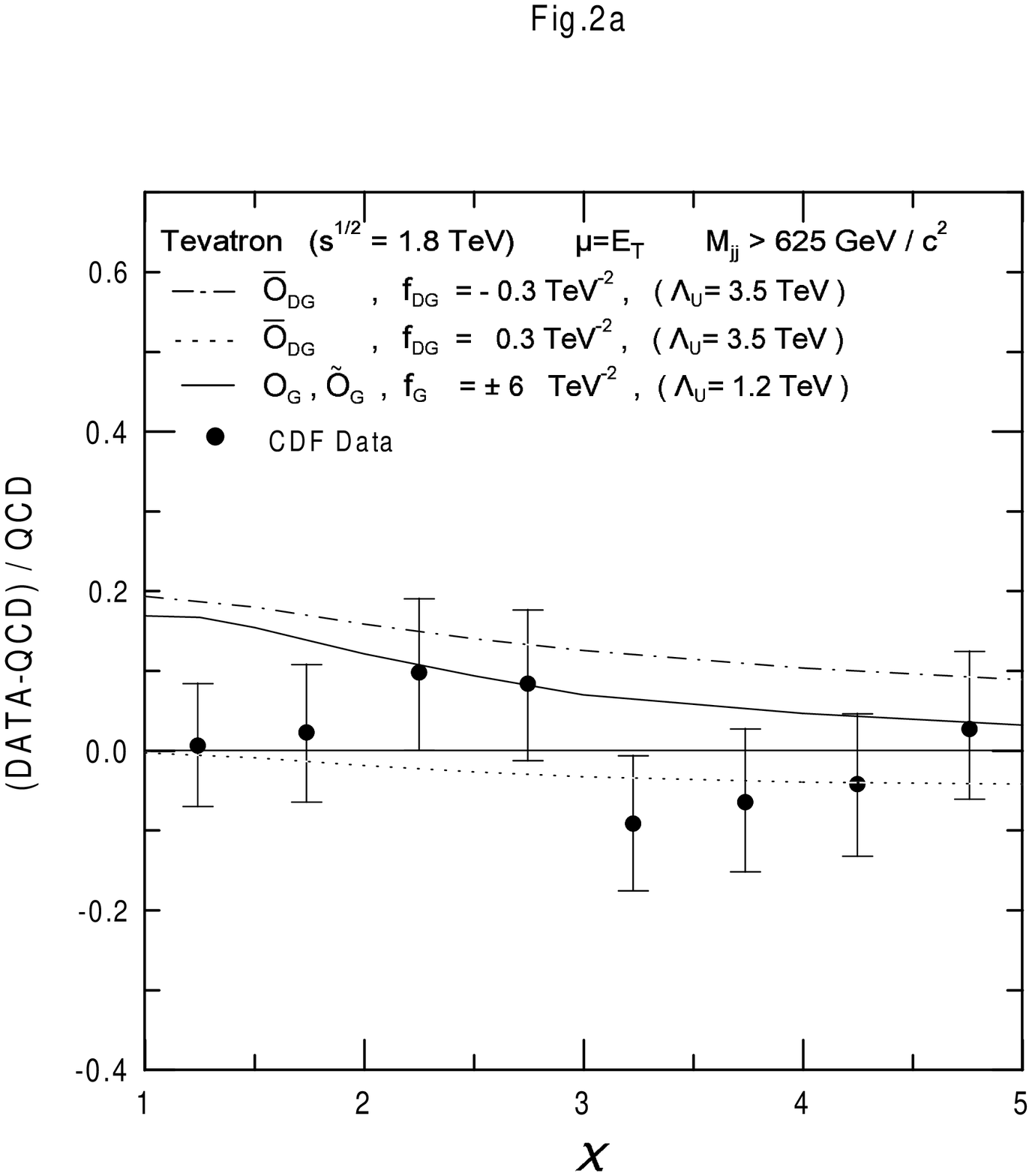,height=12cm}\hspace{0cm} 
\epsfig{file=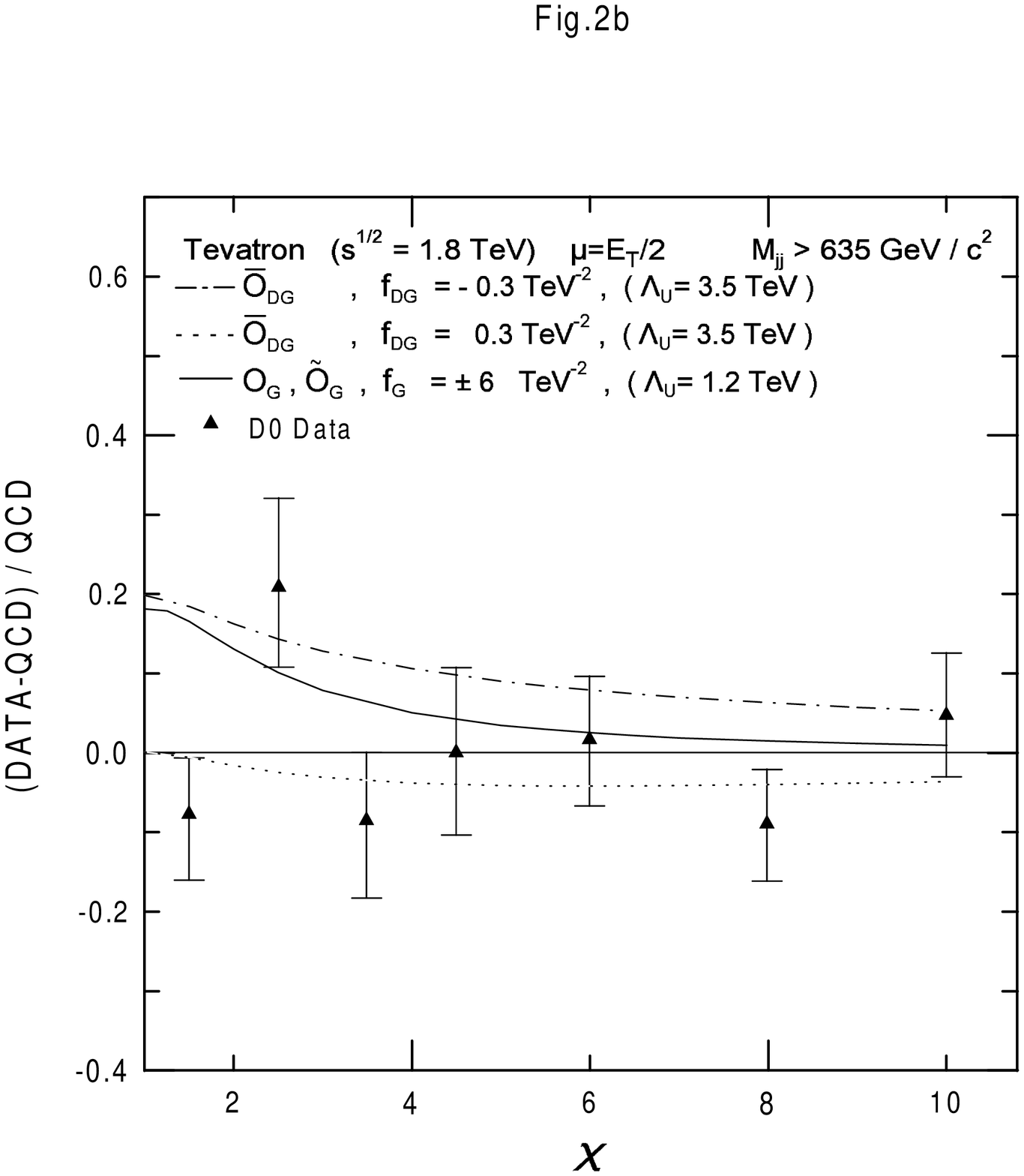,height=12cm}
\]
\vspace*{-3.5cm}\null\\
\hspace*{1.5cm} (a) \hspace{7.5cm}  (b)
\vspace*{0.5cm}
\[
\epsfig{file=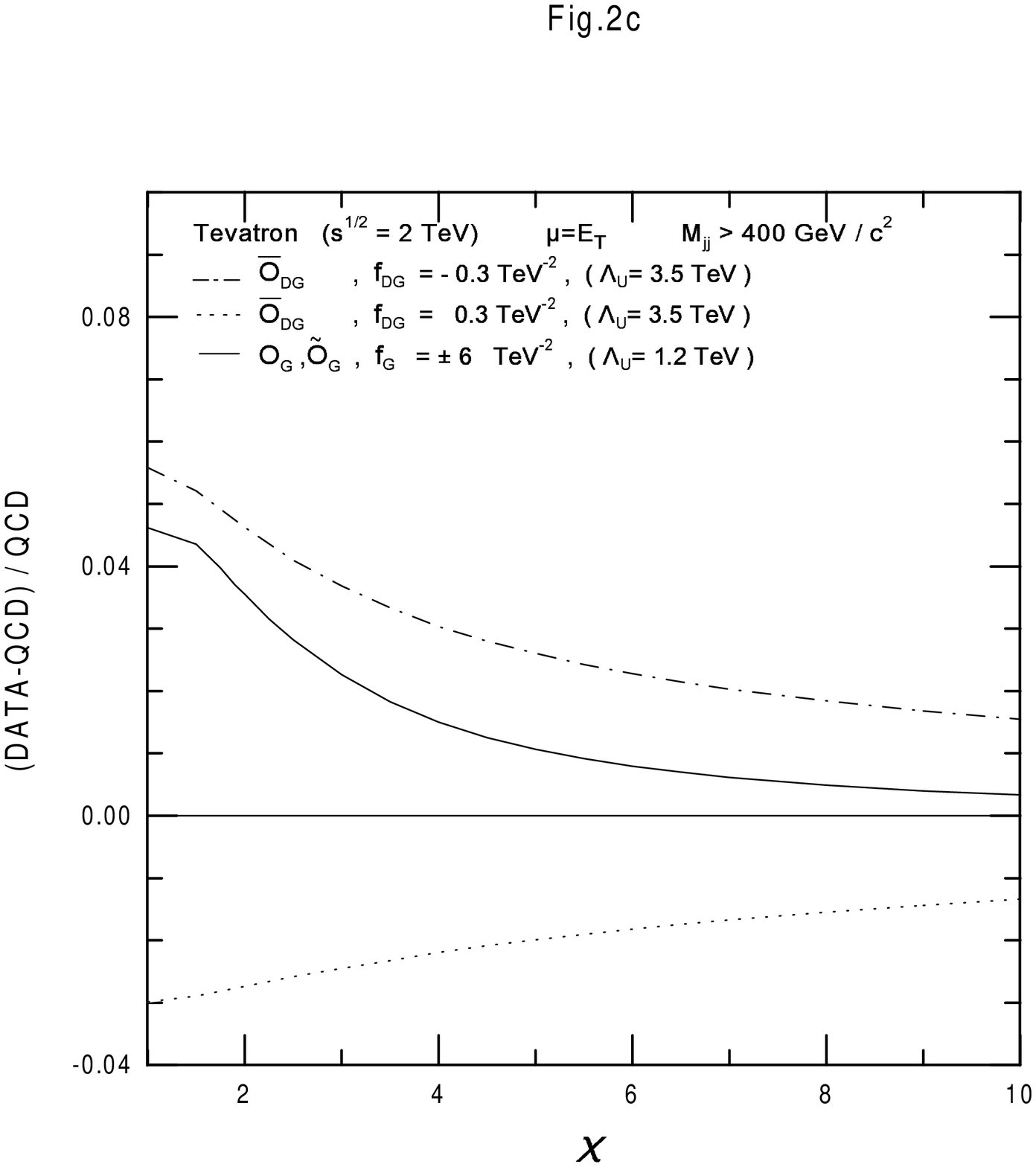,height=12cm}\hspace{0cm}
\epsfig{file=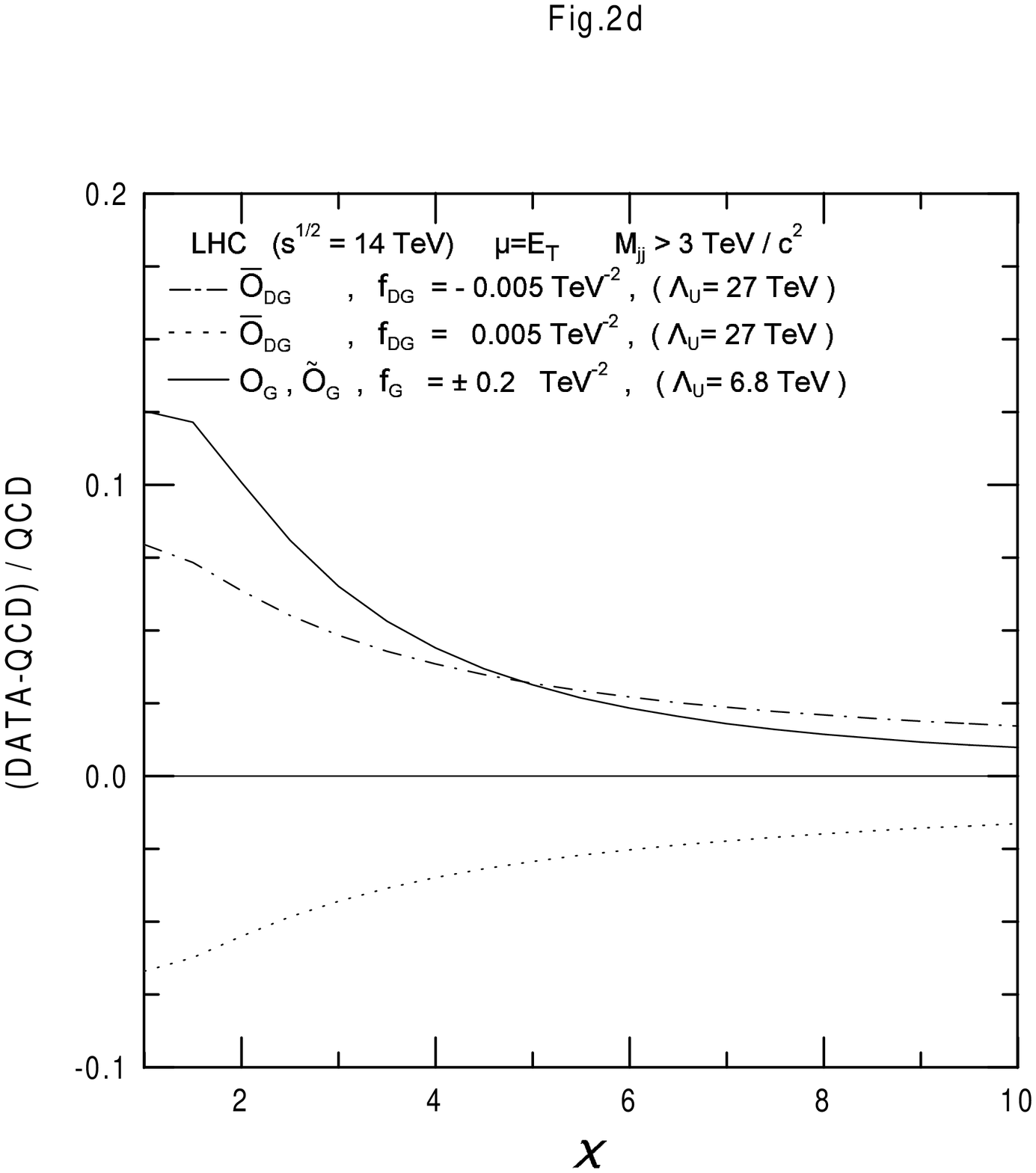,height=12cm}
\]
\vspace*{-3.5cm}\null\\
\hspace*{1.5cm} (c) \hspace{7.5cm}  (d)\\
\vspace*{1.5cm}
\caption[1]{Possible  NP contribution from $\ol{\O}_{DG}$ 
or $\O_G,~\wtil{\O}_G $ to the dijet angular distribution
compared with the CDF \cite{CDF-chi} (a) and D0 
\cite{D0-chi} (b) data, and a
possible signal at the upgraded Tevatron (c) and LHC (d). 
In the $\wtil{\O}_G$ case the
results correspond to $\wtil{f}_G$  equal to the
indicated $f_G$.}
\end{figure}

\newpage

\begin{figure}[p]
\vspace*{-5.cm}
\[
\epsfig{file=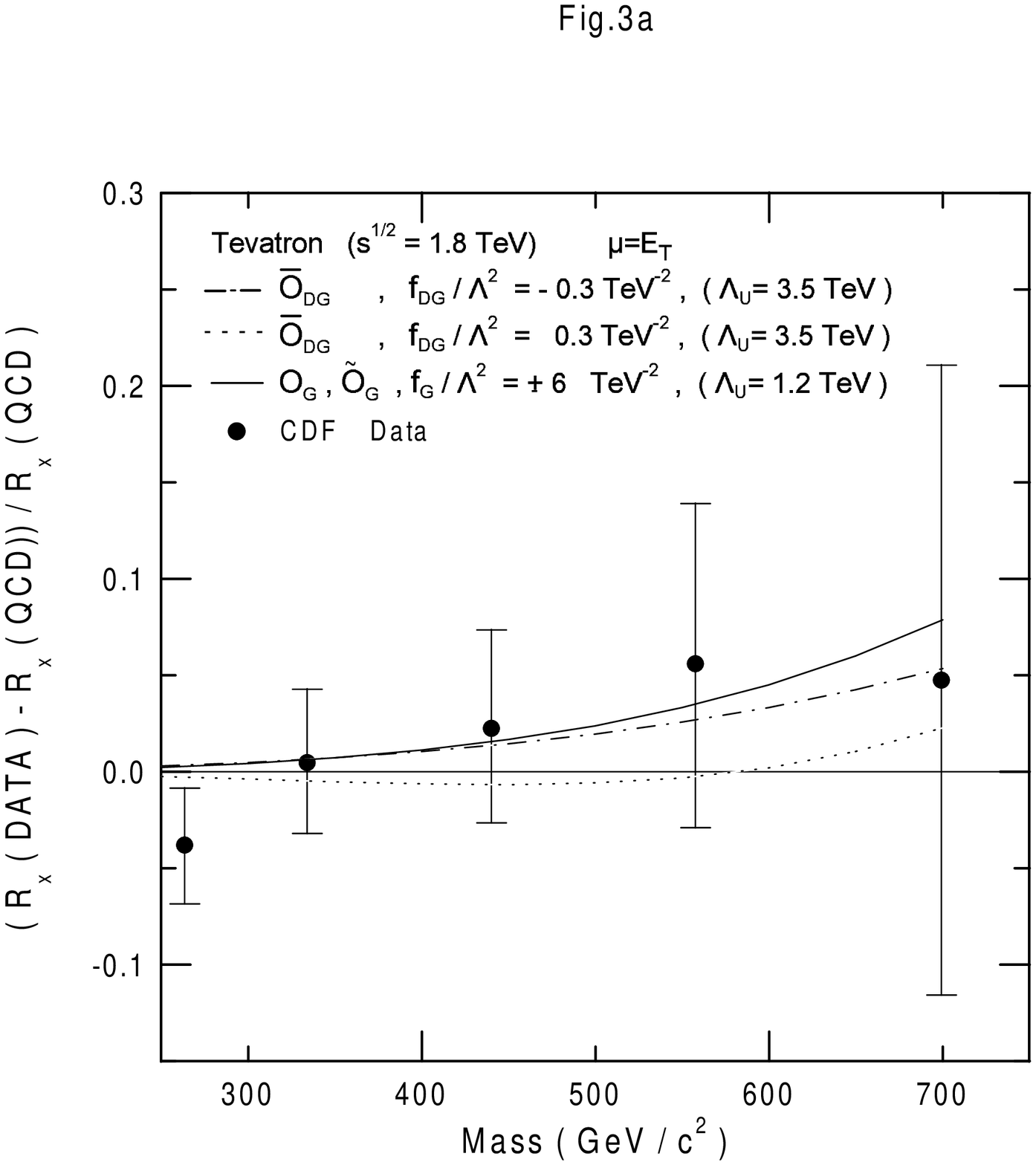,height=12cm}\hspace{0cm} 
\epsfig{file=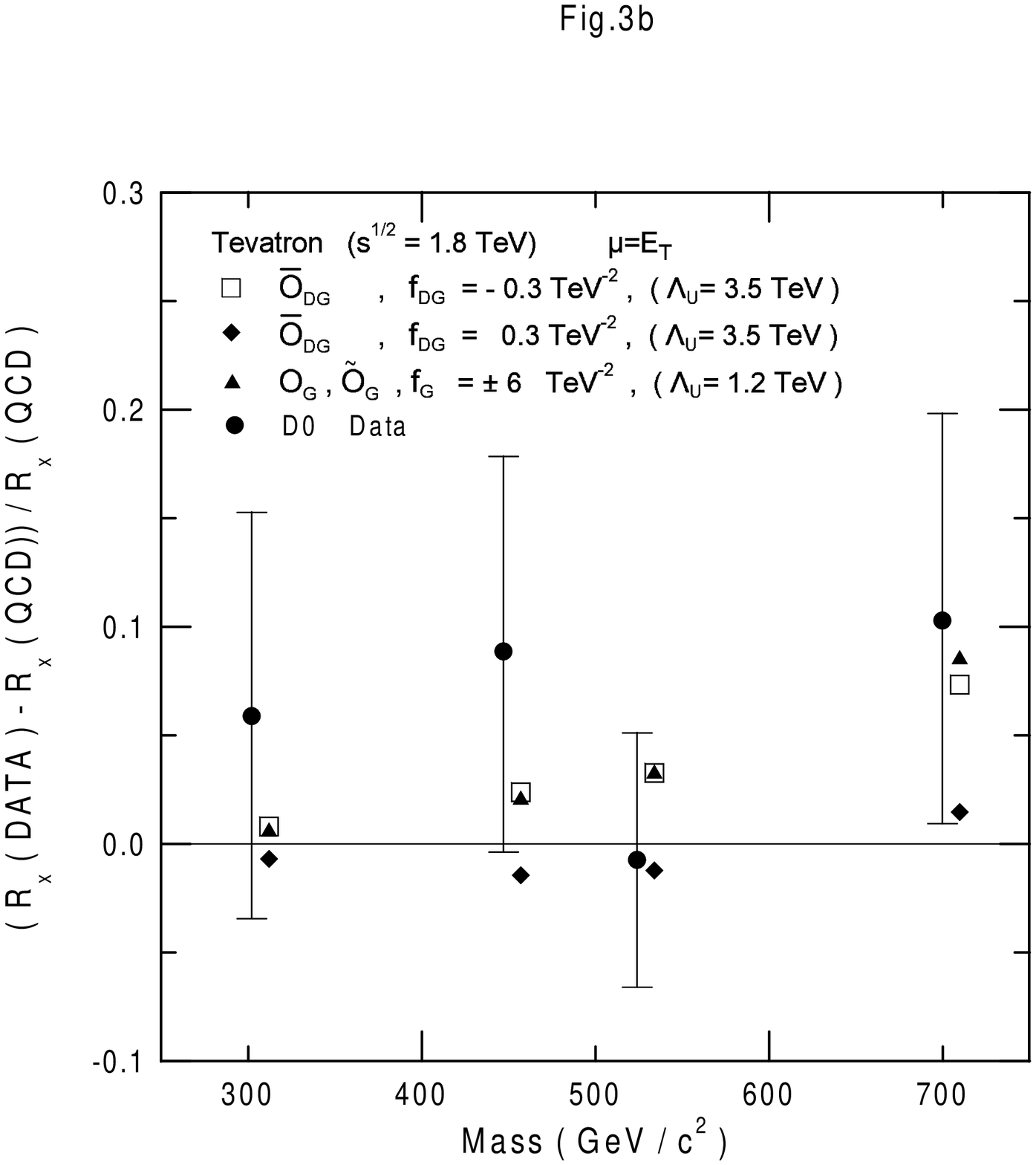,height=12cm}
\]
\vspace*{-3.5cm}\null\\
\hspace*{1.5cm} (a) \hspace{7.5cm}  (b)
\vspace*{0.5cm}
\[
\epsfig{file=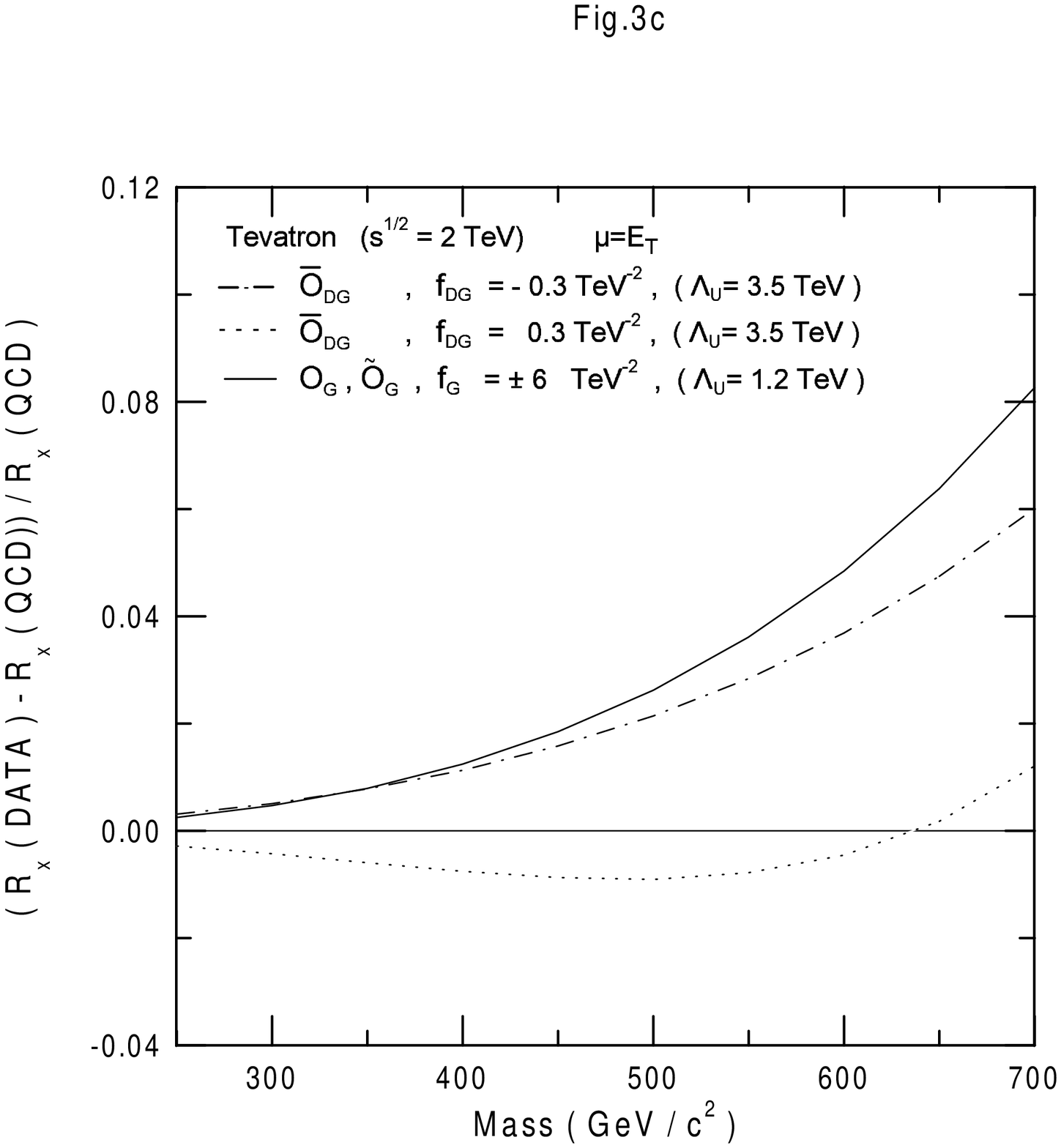,height=12cm}\hspace{0cm}
\epsfig{file=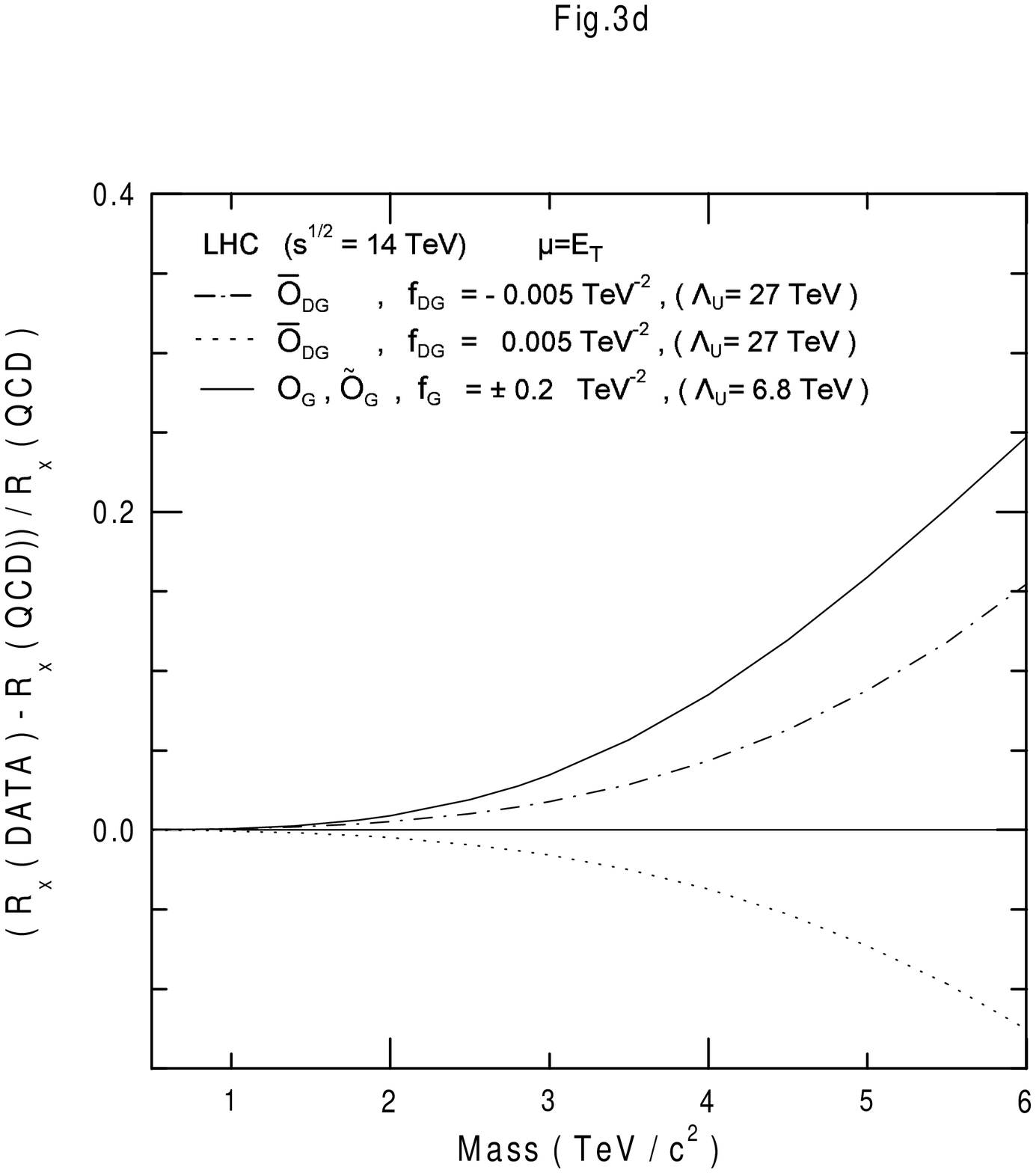,height=12cm}
\]
\vspace*{-3.5cm}\null\\
\hspace*{1.5cm} (c) \hspace{7.5cm}  (d)\\
\vspace*{1.5cm}
\caption[1]{Possible  NP contribution from $\ol{\O}_{DG}$ 
or $\O_G,~\wtil{\O}_G $ to the $R_\chi$ dependence on the 
dijet-mass compared 
with the CDF \cite{CDF-chi} (a) and D0 
\cite{D0-chi} (b) data;
the upgraded Tevatron with
$\chi_1=3.5,~ \chi_{max}=7$ (c); and  the LHC with 
$\chi_1=5,~ \chi_{max}=10 $ (d); (see text). In (b) the theoretical 
points are offset in mass, to allow distinction from 
the experimental ones. In the $\wtil{\O}_G$ case the
results correspond to $\wtil{f}_G$ equal to the
indicated $f_G$. }
\end{figure}

\clearpage
\newpage

\begin{center}

{\large \bf Figure captions}
\end{center}
\vspace{0.5cm}

{\bf Fig.1}~~Possible  NP contribution to the $E_T$ distribution
for the inclusive jet
production averaged over $0.1 \leq |\eta| \leq 0.7 $,
from $\ol{\O}_{DG}$ or $\O_G,~\wtil{\O}_G $;
compared with the CDF and D0 data at the Tevatron 
\cite{CDF-ET, D0-ET, Albrow} (a),
and a possible LHC signal (b). In the $\wtil{\O}_G$ case the
results correspond to $\wtil{f}_G$ equal to the
indicated $f_G$.\\

{\bf Fig.2}~~Possible  NP contribution from $\ol{\O}_{DG}$ 
or $\O_G,~\wtil{\O}_G $ to the dijet angular distribution
compared with the CDF \cite{CDF-chi} (a) and D0 \cite{D0-chi} 
(b) data, and a
possible signal at the upgraded Tevatron (c) and LHC (d). 
In the $\wtil{\O}_G$ case the
results correspond to $\wtil{f}_G$  equal to the
indicated $f_G$.\\

{\bf Fig.3}~~Possible  NP contribution from $\ol{\O}_{DG}$ 
or $\O_G,~\wtil{\O}_G $ to the $R_\chi$ dependence on the 
dijet-mass compared 
with the CDF \cite{CDF-chi} (a) and D0 \cite{D0-chi}
(b) data; the upgraded Tevatron with
$\chi_1=3.5,~ \chi_{max}=7$ (c); and  the LHC with 
$\chi_1=5,~ \chi_{max}=10 $ (d); (see text). In (b) the theoretical 
points are offset in mass, to allow distinction from 
the experimental ones. In the $\wtil{\O}_G$ case the
results correspond to $\wtil{f}_G$ equal to the
indicated $f_G$.

\end{document}